\input phyzzx.tex

\twelvepoint

%%%%%%%%%%%%%%%%%%%%%%Some Definitions%%%%%%%%%%%%%%%%%%%

%%%%%%%%%%%%%%%%%%%%%%%%%%%%References%%%%%%%%%%%%%%%%%%%

\REF\Strominger{A. Strominger, Phys. Lett. {\bf B383} (1996) 44,
hep-th/9512059}
\REF\DVV{R. Dijkgraaf, E Verlinde and H. Verlinde, Nucl. Phys. {\bf B486}
(1997) 77, hep-th/9603126; Nucl. Phys. {\bf B486} 89 (1996), hep-th/9604055}
\REF\Witten{E. Witten, in Strings '95: Future Perspectives in String Theory,
hep-th/9507121}
\REF\Seiberg{N. Seiberg, {\it New Theories in Six Dimensions and the
Matrix Description of M Theory on ${\bf T}^5$ and ${\bf T}^5/{\bf Z}_2$},
hep-th/9705221}
\REF\HS{P.S. Howe and E. Sezgin, Phys. Lett. {\bf B394} (1997) 62,
hep-th/9611008}
\REF\HSW{P.S. Howe, E. Sezgin and P.C. West, Phys. Lett. {\bf B399}
(1997) 49, hep-th/9702008}
\REF\S{M. Aganagic, J. Park, C. Popescu, and J. H. Schwarz,
Worldvolume action of the M-theory fivebrane,
hep-th/9701166,
I. Bandos, K Lechner, A. Nurmagambetov, P. Pasti and D. Sorokin, and M.
Tonin, Covariant action for the super fivebrane of M-theory,
hep-th/9701149}
\REF\PS{M. Perry and J.H. Schwarz, Nucl. Phys. {\bf B489} (1997) 47,
hep-th/9611065}
\REF\SEZ{ P.S. Howe, E. Sezgin and P.C. West, Phys. Lett.{\bf B400 }
(1997) 255}
\REF\Verlinde{E. Verlinde, Nucl. Phys. {\bf B455} (1995) 211,
hep-th/9506011}
\REF\GLPT{M.B. Green, N.D. Lambert, G. Papadopoulos and P.K. Townsend,
Phys. Lett. {\bf B384} (1996) 86, hep-th/9605146}
\REF\DL{M.J. Duff and J.X. Lu, Nucl. Phys. {\bf B416} 301;
Phys. Lett. {\bf B273} (1991) 409}
\REF\CM{C.G. Callan and J.M. Maldecena, {\it Brane Death and Dynamics From
the Born-Infeld Action}, hep-th/9708147}
\REF\Gibbons{G.W. Gibbons, {\it Born-Infeld Particles and Dirichlet 
$p$-branes}, hep-th/9709027}

%%%%%%%%%%%%%%%%%%%%%%%%%%%%Title Page%%%%%%%%%%%%%%%%%%%
\pubnum={KCL-TH-97-51\cr hep-th/9709014}
\date{August 1997}

\titlepage

\title{\bf The Self-Dual String Soliton}

\centerline{P.S. Howe}
\centerline{N.D. Lambert}

\centerline{and}

\centerline{P.C. West\foot{phowe, lambert, pwest@mth.kcl.ac.uk}}
\address{Department of Mathematics\break
         King's College, London\break
         England\break
         WC2R 2LS\break
         }

\abstract
We obtain a BPS soliton of the M theory fivebrane's
equations of motion representing a supersymmetric self-dual string. The
resulting solution  is then  dimensionally reduced and used
to obtain $0$-brane and $(p-2)$-brane solitons on D-$p$-branes.

\endpage

%%%%%%%%%%%%%%%%%%%%%%%%%%%%%%%%Chapter One%%%%%%%%%%%%%%%

\chapter{Introduction}

In recent years it has become clear that the still rather mysterious
M theory  governs many aspects of the lower dimensional string
theories. What little is known  of M theory is powerful enough to lead us to
new phenomena in string theory and indeed new string theories. In particular
the M theory
fivebrane  is strongly
believed hold a new kind of self-dual string theory on its worldvolume
[\Strominger,\DVV].
This new and somewhat elusive theory
has also  appeared in other contexts such as
the compactification of type IIB
string theory on $K3$ [\Witten],
M(atrix) theory on ${\bf T}^5$ [\Seiberg] and the S-duality of
$N=4$, $D=4$ super-Yang-Mills [\Witten]. Thus one may hope that a greater
understanding
of this self-dual string may lead directly to a greater understanding
of duality, string theory and M theory.

In this paper we shall seek a supersymmetric string soliton solution
to the M theory fivebrane's equations of motion [\HS,\HSW]. We will
use the six-dimensional covariant  field equations of motion
derived in [\HSW]. Alternative formulations of the fivebrane were given in
[\S]. In [\PS] a smooth string soliton solution of the field equations
for an interacting second rank tensor gauge field was found.
 However, it is known [\SEZ] that the  fivebrane equations of
motion of [\HSW] reduce to those of [\PS] when only the
second rank tensor gauge field  is active and as a result  the
solution of [\PS] can be lifted to be a solution of the full fivebrane
equations of motion. However
the solution so obtained  takes all the scalar fields to be
constant  and it cannot preserve any supersymmetries since there is
nothing to cancel off  the tensor field  force (this will also be
apparent from the supersymmetry  transformation given below).  Instead
it represents a non-singular field configuration around a string,
analogous to the Born-Infeld expression for the electric field of a
charged point particle [\PS]. In the next section we recall the
fivebrane equations of motion and derive the supersymmetry
transformation of the fermion field. From this result we 
derive the fivebrane analogue of the Bogomol'nyi conditions which
preserve half the supersymmetry. In section three we  use this
Bogomol'nyi condition to  solve the field equations and find the
solution corresponding to the self-dual string.
In section four we  dimensionally reduce the string soliton to
obtain solitons of the type II D-$p$-brane worldvolume theories.  In
the final section we  conclude with some comments on our solutions.

%%%%%%%%%%%%%%%%%%%%%%%%%%%%%%%%Chapter Two%%%%%%%%%%%%%%%

\chapter{The Fivebrane and Its Worldvolume Supersymmetry}

We use the fivebrane equations and conventions of
[\HSW]. In this paper the fivebrane is embedded in flat eleven-dimensional 
Minkowski superspace. We must distinguish between world
and tangent indices, fermionic and bosonic indices  and
indices associated with the target space
$\underline M$ and the fivebrane worldvolume $M$ . On the fivebrane
worldvolume   the bosonic tangent space indices are denoted by
$a,b,...= 0,1,2,...,5$ and bosonic world indices by
$m,n,...=0,1,2,...,5$. For example,  the inverse
vielbein of the bosonic sector of
 the fivebrane worldvolume is denoted by
$E_a^{\ m}$. The bosonic indices of the tangent space of the
target space $\underline M$ are denoted by the same symbols, but
underlined, i.e. the inverse vielbein in the bosonic sector is given
by
$E_{\underline a}^{\ \underline m}$. The fermionic indices follow the
same pattern,  those in the tangent space are denoted by $\alpha $
and $\underline \alpha $ for worldvolume $M$ and target space
$\underline M$
respectively, while the world spinor indices are denoted by $\mu$ and
$\underline \mu$.
\par
The fivebrane sweeps out a superspace $M$ in the target superspace
$\underline M$
which is specified in local coordinates
$Z^{\underline M}= (X^{\underline m}, \Theta ^{\underline \mu})
,\ \underline m=0,1,\dots 10,\ \underline \mu=1,\dots,32$.
These coordinates are functions of the worldvolume superspace
parameterised by $z^{ M}= (x^{ m}, \theta^{ \mu}),\  m=0,1,\ldots
,5;\ \mu = 1,\ldots ,16$. The $\theta^{ \mu}$ expansion
of the $z^{\underline M}$ contains
$x^m$ dependent fields of which the only independent ones  are their
$\theta^\mu=0$ components, also denoted
$X^{\underline m}$ and $ \Theta ^{\underline \mu}$, and a
self-dual tensor $h_{abc}$ which occurs as at level $\theta ^\mu $
in $ \Theta ^{\underline \mu}$. Despite the redundancy of notation
it  will be clear from the context when we are discussing the component
fields and the superfields.
\par
The bosonic target space indices of $\underline M$ may be decomposed as
those that lie in the  fivebrane worldvolume and those that lie in the
space transverse to the fivebrane; we denote these indices by $a$ and
$a'$ respectively  (i.e. $\underline a=(a,a'), a=0,1,\ldots,5;\
a'=1',\ldots,5'$) with a similar convention for  world indices. The
initially thirty-two component spinor indices $\underline \alpha$ are
split into a pair of sixteen component spinor indices (i.e.
$\underline \alpha=(\alpha,
\alpha'),\alpha=1,\ldots,16;\ \alpha'=1',\ldots,16'$) corresponding to
the breaking of half of the supersymmetries by the fivebrane.
\par
We will use the super-reparameterisations of the
worldvolume to choose the so called  static
gauge. In this gauge we identify the bosonic  coordinates in
the worldvolume with the bosonic   coordinates on  the worldvolume
(i.e. $ X^n=x^n,\ n=0,1,\ldots ,5$) and set the fermionic fields
$\Theta ^\alpha=0, \  \alpha=1,\ldots,16$. For a flat background
$\Theta ^{\underline \mu}=\Theta ^{\underline \alpha}
\delta _{\underline \alpha}^{\underline \mu}$. Thus  the component
field content of the fivebrane is
$X^{a'}$ $(a'=1',\dots,5')$, $\Theta ^{\alpha'}$ $(\alpha'=1',\ldots,16')$ 
and the self-dual field strength
$h_{abc}$. These fields contribute 5, 8 and 3 on-shell degrees of
freedom  respectively and  belong to the tensor multiplet of
six-dimensional $(2,0)$ supersymmetry. We will discuss below the
decomposition of the eleven-dimensional Lorentz and spin groups with respect  
to the fivebrane and  its consequences for the spinor index notation.
\par
We are interested in solutions to the field equations in a flat
background and so are required to solve the equations when
$\Theta^{\alpha'}=0$. In this limit the bosonic equations of the
five brane are given by [\HSW]
$$\eqalign{
G^{mn}\nabla_m\nabla_nX^{a'}=0\ ,\cr
G^{mn}\nabla_m H_{npq}=0\ .\cr}
\eqn\fieldeq
$$
The field tensor $H_{npq}$ is related to $h_{abc}$ by
$H_{mnp}=E_m^{\ a}E_n^{\ b}E_m^{\ c}m_{b}^{\ d}m_{c}^{\ e}h_{ade}$
where $m_a^{\ b} \equiv \delta_a^{\ b} - 2h_{acd}h^{bcd}$. Although
$H_{npq}$ is not self-dual, it does  obey the usual Bianchi identity
and, as a consequence, is the curl of  the two form gauge field
$B_{mn}$ (i.e.
$H_{npq}= \partial _{[n}B_{pq]})$. This is  in contrast to $h_{abc}$,
which is self-dual but which is not in general 
the curl of a gauge field. As such, we
must in addition to solving \fieldeq\ ensure that $H_{mnp}$ is closed
 and that the corresponding $h_{abc}$ is self-dual.
 The worldvolume   vielbein $E_m^{\ a}$ is specified by  the relation
$E_m^{\ a}= e_m^{\ b} {(m^{-1})}_b^{\ a}$, where the
vielbein $e_m^{\ a}$ is that more usually associated with
branes and is defined  by
$$ 
e_m^{\ a} \eta_{ab} e_{n}^{\ b}= g_{mn}=
\partial_n Z^{\underline N} \partial_m Z^{\underline M}
E_{\underline M}^{\ \ {\underline a}}
\eta_{\underline{ab}}E_{\underline N}^{\ \ {\underline b}}\ .
\eqn\gsmeq
$$
In equation \fieldeq\ $G_{mn}= E_m^{\ a} E_n^{\ b} \eta _{ab}$ and
 the covariant derivative $\nabla_m$ is constructed from the
Levi-Civita connection corresponding to  the
 metric $g_{mn}$.
\par
In addition to solving the fivebrane field equations we wish to
find a solution which preserves half the supersymmetry. As a result,
we are required to find the supersymmetry transformation of
$\Theta ^{\alpha'}$ and then the fivebrane analogue of the Bogomol'nyi
equation. To this end, we  recall some of the salient points
of the super-embedding formalism. The frame vector fields on the target
manifold $\underline M$ and the fivebrane worldvolume submanifold
$M$ are given by $E_{\underline A} = E_{\underline A} ^{\ \underline M}
\partial_{\underline M}$ and
$E_{A} = E_{A} ^{\ M} \partial_{M}$ respectively.
The coefficients $E_{A}^{\ \underline A}$ encode the relationship
between the vector fields $E_{A}$ and $E_{\underline A}$, i.e. $E_{A}
= E_{A}^{\ \underline A}E_{\underline A}$. Applying this relationship
to the coordinate $z^M$ we find the equation
$$
E_{A}^{\ \underline A}= E_{A}^{\ \ N}\partial_N Z^{\underline M}
E^{\ \underline A}_{\underline M}\ .
\eqn\ez
$$
\par
It can be shown that the vector fields on the fivebrane can be
chosen such  that
$$
E_{\alpha}^{\ \underline \beta} = u_{\alpha}^{\ \underline \beta} +
h_{\alpha}^{\ \gamma'}u_{\gamma'}^{\ \underline \beta}, \ \ 
 E_{\alpha'}^{\ \underline \beta} = u_{\alpha'}^{\ \underline \beta}\ ,
\eqn\Edef
$$
and
$$ E_a^{\ \underline a}= m_a^{\ b} u_b^{\ \underline a},\ \ 
E_{a'}^{\ \underline a}= u_{a'}^{\ \underline a}\ .
\eqn\edeftwo
$$
In equation \Edef\ the tensor $h_{\alpha}^{\ \beta'}$ is related to
the  three-form $h_{abc}$  [\HSW] by
$$
h_{\alpha}^{\ \beta'}\rightarrow h_{\alpha i \beta}^{\ \ \ j} =
{1\over 6}\delta_{i}^{\ j}(\gamma^{abc})_{\alpha\beta}h_{abc} \ .
\eqn\h
$$
The matrix $u_{\overline a}^{\ \underline b}
\equiv (u_{ \overline a}^{\  b},u_{\overline a}^{\
 b'})$ is an  element of $SO(1,10)$ and the matrix
$u_{\overline \alpha}^{\ \underline \beta}\equiv
(u_{\overline \alpha}^{\
\beta},u_{\overline \alpha'}^{ \  \beta'})$ forms an element
of
$Spin(1,10)$. As is clear from the notation, the indices with an
overbar take the same  range as those with  an underline. We
recall that the connnection between the Lorentz and spin groups   is
given by
$$ u_{\overline \alpha}^{\ \underline \gamma} u_{\overline
\beta}^{\ \underline \delta}
{(\Gamma^{\underline a})}_{\underline \gamma \underline \delta }
={(\Gamma^{\overline b})}_{\overline \alpha \overline\beta }
u_{\overline b}^{\ \underline a}\ .
\eqn\spin
$$
\par
In the presence of a flat superspace target space the
super-reparameterisation invariance reduces to translations and rigid
supersymmetry transformations. The latter take the form
$$
\delta x^{\underline n}= {i\over2}\Theta \Gamma^{\underline n}
\epsilon\ ,\ \ 
\delta \Theta ^{\underline \mu}= \epsilon^{\underline \mu}\ .
\eqn\rs
$$
Unlike other formulations, the super-embedding approach of
[\HS,\HSW] is invariant under super-reparameterisations of the
worldvolume, that is, invariant under
$$
\delta z^{ M} = -v^{ M} \ ,
\eqn\superdiff
$$
where $v^{ M}$ is a supervector field on the fivebrane worldvolume.
The corresponding motion induced on the target space $\underline M$
is given by
$$ \delta Z^{\underline B} = v^A E_A^{\ \underline B}\ ,
\eqn\va
$$
where $v^M= v^AE_A^{\ M}$ and rather than use the embedding
coordinates $Z^{\underline N}$ we  refered them to the background
tangent space, i.e.
$Z^{\underline B}\equiv Z^{\underline M} E_{\underline M}^{\ \ \underline B}$. 
We are interested  in supersymmetry transformations and
so consider
$v^a=0,\ v^\alpha
\not=0$; with this choice and including the rigid supersymmetry
transformation of the target space of equation \rs\ the
transformation of $\Theta^{\underline \alpha}$ is
given by [\HSW]
$$
\delta\Theta^{\underline \alpha} = v^{\beta}
E_{\beta}^{\ \underline \alpha}
+ \epsilon^{\underline\alpha}\ .
\eqn\odddiff
$$
The local supersymmetry tranformations $v^{\alpha}$ are used to
set $\Theta^{\alpha}=0$ which is part of the static gauge choice.
However, by combining these transformations with those of the rigid
supersymmetry  of the target space $\epsilon ^{\alpha}$ we find a
residual rigid  worldvolume supersymmetry which is determined by the
requirement  that the gauge choice $\Theta^{\alpha}=0$ is preserved.
Consequently,  we require $v^{\beta} = -\epsilon^{\alpha}(E^{-1})^{\
\beta}_{\alpha}$.
We then find the supersymmetry transformation for the fermions
is given by
$$
\delta\Theta^{\alpha'} = \epsilon^{\alpha}
(E^{-1})^{\ \beta}_{\alpha}E_{\beta}^{\ \alpha'}\ ,
\eqn\susyfull
$$
where we have chosen   $\epsilon^{\alpha'}=0$.
\par
To evaluate this expression we are required to find
$E_{A}^{\ \underline A}$, or equivalently the $u$'s of $SO(1,10)$ and
$Spin(1,10)$, in terms of the component fields in the limit
$\Theta^{\underline \alpha }=0$.  Using equation \ez\ , the Lorentz
condition $u_c^{\ \underline a} \eta
_{\underline a \underline b} u_d^{\ \underline b}=\eta_{cd}$  and the
static gauge choice
$X^n=x^n$ we find that
$$(u_a^{\ b},u_a^{\ b'})=(e_a^{\ n} \delta _n^b,e_a^{\ n}\partial _n
x^{b'} )\ .
\eqn\lu
$$
Using the remaining Lorentz conditions we find,
up to a local  $SO(5)$ rotation, that the full lorentz matrix
$u_{\overline a}^{\ \underline a}$ is given by
$$u =\left(\matrix{
e^{-1}                              &e^{-1}\partial X \cr
-d^{-1}{(\partial X)}^T {(\eta_1)}^T&d^{-1} \cr }\right)\ ,
\eqn\um
$$
where the matrix $d$ is defined by the condition
$dd^T= I+ {(\partial X)}^T \eta_1 {(\partial X)}$,
${(\partial X)}^T $ is the transpose of the matrix ${(\partial_n
X^{a'})}$  and $\eta_1$ is the Minkowski metric on the fivebrane and is
given by
$\eta_1 = diag (-1,1,1,1,1,1)$.
\par
The $u_{\overline \alpha}^{\ \underline \beta}\in \ Spin(1,10)$ corresponding
to the above
$u_{\overline a}^{\ \underline b}\in \ SO(1,10)$ are found using  
equation \spin\ .
It is instructive to carry out this calculation at the linearized
level. To this order
$u^{\ \beta}_{\alpha}=\delta^{\ \beta}_{\alpha}$,
$u_{\alpha'}^{\ \beta'}
=\delta_{\alpha'}^{\ \beta'}$ while
$u^{\ \beta}_{\alpha'}$ and
$u_{\alpha}^{\ \beta'}$ are linear in the fields.
Taking $(\overline \alpha, \overline \beta) =(\alpha,  \beta)$
equation \spin\  becomes
$$ u_{\alpha}^{\gamma'}{(\Gamma^{d'})}_{\gamma' \beta}
+ (\alpha \leftrightarrow\beta )= {(\Gamma^{a})}_{\alpha \beta}
\partial X^{d'}\ .
\eqn\ul
$$
\par
In order to  analyse this equation we discuss the decomposition of the
spinor indices in more detail. We recall that the bosonic  indices of
the fields  on the fivebrane can be decomposed into
longitudinal and transverse indices i.e. $\underline a=
(a,a')$ according to the decomposition of the Lorentz group
$SO(1,10)$ into
$SO(1,5)\times SO(5)$. The corresponding decomposition of the
spin groups is $Spin (1,10)\to\ Spin(1,5) \times USp(4)$. The spinor
indices of the groups $Spin(1,5)$ and  $ USp(4)$ are denoted
by  $\alpha,\beta,...=1,...,4$ and $i,j,...=
1,...,4$  respectively. Six-dimensional Dirac spinor indices normally take
eight values, however the spinor indices we use for $Spin(1,5)$
correspond to  Weyl spinors. Although we began with spinor
indices $\underline \alpha $ that took  thirty-two dimensional values
 and were broken into two pairs of indices each taking sixteen values
${\underline
\alpha}=(\alpha,\alpha')$, in the final six-dimensional
expressions the spinor indices  are further decomposed according
to the above decomposition of the spin groups and we take
$\alpha\rightarrow\alpha i$
and $\alpha'\rightarrow{}_{\alpha}^{i}$ when appearing as
superscripts and $\alpha\rightarrow\alpha i$
and $\alpha'\rightarrow{}^{\alpha}_{i}$ when appear as subscripts
[\HSW].
It should be clear whether we mean $\alpha$ to be sixteen or
four dimensional
depending on the absence or presence of $i,j,...$ indices respectively.
For example, we will write  $\Theta ^{\alpha'}\to\Theta_{\alpha}^{i}$.
\par
Using the corresponding decomposition of the spinor indices and the
expressions
$$ (\Gamma^{a'})_{\alpha' \beta}=- (\gamma^{a'})_{ij}
\delta^{\alpha}_{\beta},\
(\Gamma^a)_{\alpha \beta}=- (\gamma^a)_{\alpha \beta}
\eta_{ij}\ ,
\eqn\ga
$$
where $\eta_{ij}$ is the anti-symmetric invariant tensor of $USp(4)$
in  equation
\ul\ we find that
$$
u_{\alpha}^{\ \beta'}\rightarrow u_{\alpha i\beta}^{\ \ \ j}
= -{1\over2} (\gamma^a)_{\alpha\beta}(\gamma_{\underline b'})_{i}^{\ j}
\partial_a X^{\underline b'} \ .
\eqn\u
$$
From equation \odddiff\ we find that the linearised
supersymetry transformation is given by
$$
\delta_0\Theta_{\beta}^{\ j} = \epsilon^{\alpha i}\left(
{1\over2}(\gamma^a)_{\alpha\beta}(\gamma_{\underline b'})_{i}^{\ j}
\partial_a X^{\underline b'}
- {1\over6}(\gamma^{abc})_{\alpha\beta}\delta_i^{\ j}h_{abc}
\right)\ .
\eqn\susy
$$
\par
We now wish to find the fivebrane analogue of the Bogomol'nyi
condition. We will consider a static fivebrane whose world
sheet lies in the $x^0,x^1$ directions and take all fields to be
independent of $x^0$ and $x^1$. We take the transverse
scalars $X^{a'},\ a'=1',2',\ldots,5'$ to lie only in the five direction.
Denoting $X^{5'}=\phi$ and $h_{01a} = v_{a},\ a=2,\ldots,5$ and taking
$h_{0ab}=0=h_{1ab},\ a,b=2,\ldots,5$ we find that
 the  preserved supersymmetries can be deduced from
 equation \susy\ to be
$$
\delta_0\Theta_{\alpha}^{j} =0= \epsilon^{\alpha i}\left(
{1\over4}\delta_{\alpha}^{\ \gamma}\delta_i^{\ k}\partial_{a}\phi -
(\gamma^{01})_{\alpha}^{\ \gamma}(\gamma_{ 5'})_{i}^{\
k}v_a\right)\ .
\eqn\bogomolnyi
$$
Therefore if we take
$$
V_{n}\equiv H_{01n} = \pm{1\over 4}\partial_{n}\phi ,\ \ n=2,\ldots,5\ ,
\eqn\vdef
$$
the solution will be invariant under linearised supersymmetries
which satisfy
$$
\epsilon_0^{\beta j} = \pm(\gamma^{01})_{\alpha}^{\ \beta}
(\gamma_{ 5})_i^{\ j}
\epsilon_0^{\alpha i} \ .
\eqn\ressusy
$$
At the linearised level $h_{abc}$ and $H_{mnp}\delta_a^m \delta_b^n
\delta_c^p$ are equal, however, we have taken the opportunity to write
the Bogomol'nyi equation in such a way that it will be valid for the
full supersymmetry transformations.
\par
Finally, it remains to show that the full supersymmetry
transformation
\susyfull\ vanishes if \vdef\ and \ressusy\ are satisfied. We note
that the full supersymmetry variation of $\Theta^{\alpha '}$ of
equation \susyfull\ contains the the matrix
$(E^{-1})^{\ \beta}_{\alpha}E_{\beta}^{\ \alpha'}$. Supersymmetries
 are preserved when the determinant of this matrix vanishes. This
will be the case if the determinant of the matrix
$E_{\alpha}^{\ \beta'}{(u^{-1})}_{\beta'}^{\ \gamma'}$ vanishes and,
since it is easier, we calulate this latter matrix. We find that
$$
E_{\alpha}^{\ \beta'}{(u^{-1})}_{\beta'}^{\ \gamma'}=
u_{\alpha}^{\ \beta'}{(u^{-1})}_{\beta'}^{\ \gamma'}
+h_{\alpha}^{\ \delta'}
u_{\delta'}^{\ \beta'}{(u^{-1})}_{\beta'}^{\ \gamma'}\ .
\eqn\mat
$$
\par
To illustrate how to evaluate this expression we consider the first
term which we may write as
$$u_{\alpha}^{\ \beta'}{(u^{-1})}_{\beta'}^{\ \gamma'}
={1\over 2}u_{\alpha}^{\ \underline
\beta}{(1-\Gamma_7)}_{\underline
\beta}^{\ \underline \delta}{(u^{-1})}_{\underline \delta}^{\ \gamma'}
=-{1\over 2}u_{\alpha}^{\underline
\beta}{(\Gamma_7)}_{\underline
\beta}^{\ \underline \delta}{(u^{-1})}_{\underline \delta}^{\ \gamma'}\ .
\eqn\tone
$$
Writting $\Gamma_7=
{1\over 6!}\epsilon^{a_1a_2a_3a_4a_5a_6}\Gamma_{a_1a_2a_3a_4a_5a_6}$
and using equation \spin\ we find that the first term becomes
$$-{1\over 2.6!}
\epsilon^{a_1a_2a_3a_4a_5a_6}
u_{a_ 1}^{\ \underline b_1} u_{a_2 }^{\ \underline b_2}
u_{a_ 3}^{\ \underline b_3} u_{a_ 4}^{\ \underline b_4}
u_{a_ 5}^{\ \underline b_5} u_{a_6 }^{\ \underline b_6}
{(\Gamma_{\ \underline b_1
\underline b_2 \underline b_3\underline
b_4 \underline b_5\underline b_6})}_{\alpha}^{\ \gamma'}\ .
\eqn\ttwo
$$
The final $\Gamma$-matrix in this last equation vanishes
 if  the
$\underline b_i$  indices take values in the transverse direction an
even number of  times. However, for the field configuration of interest
to us,  it also vanishes if their are two or more of these indices in
the transverse space. Implementing this restriction the first term can
be written as
$$-{1\over 2} \det(e^{-1})\partial _c X^{b'}{(\gamma^c)}_{\alpha\beta}
{(\gamma^{b'})}_i^{\ k}\ . 
\eqn\firstterm
$$
We have used the relation
${(\Gamma _{b_1\ldots b_5 b'})}_{\alpha}^{\gamma'}
=-{(\gamma _{b_1\ldots b_5 })}_{\alpha \gamma}
{(\gamma _{ b'})}_{i}^{\ k} $ and the fact that
${(\gamma^{abc})}_{\alpha \beta}$ is anti-self-dual on its vector
indices to carry out this last step .
\par
Evaluating the second term of equation in a similar manner we find
that the matrix of equation \mat\
becomes
$$E_{\alpha}^{\ \beta'}{(u^{-1})}_{\beta'}^{\ \gamma'}=
-{1\over 2} \det(e^{-1})\partial _n
\delta^n_c X^{b'}{(\gamma^c)}_{\alpha\beta}  {(\gamma_{b'})}_i^{\ k}
+ {1\over 12}(1+ \det(e^{-1}))h_{abc} {(\gamma^{abc})}_{\alpha \beta}
\delta_i^k\ .
\eqn\mattwo
$$
\par
Substituting in the field configuration discussed above we find that
the Bogomol'nyi condition for the full non-linear supersymmetry is
given by
$$v_c ={1\over 2} {1\over 1+ \det(e) }\delta^n_c
\partial _n \phi\ .\
\eqn\bognon
$$
The preserved supersymmetries have a parameter $\epsilon^\alpha$ which
satisfies  equation  \ressusy\ provided we replace
$\epsilon_0^{\alpha}$in that equation  by
$\epsilon^{\alpha} E^{\ \beta}_{\alpha}$.   We will shortly
show that this condition is the same  as that of equation \bogomolnyi\
.

%%%%%%%%%%%%%%%%%%%%%%%%%%%%%%%%Chapter Three%%%%%%%%%%%%%%%

\chapter{The Self-Dual String as a Soliton}

Let us look for a string soliton
whose worldsheet lies in the $x^0,x^1$ plane. We take all fields to be
independent  of $x^0$ and $x^1$. In this section, we
denote the six-dimensional worldvolume indices which take the full
range with a hat, i.e. $\hat a,\hat b,...,\hat m,\hat n... =0,1,\ldots,5$,
and denote the  four  coordinates transverse to $x^0,x^1$ as
$a,b,...,m,n... = 2,3,4,5$.    We
take only one of the scalar fields $X^{a'}$ to be active thus
breaking the  $SO(5)$ symmetry to $SO(4)$. We choose this scalar
field to be $X^{5'}$,  the other scalar fields being constants.
Consider the ansatz
$$\eqalign{
X^{ 5'} &= \phi \ ,\cr
h_{01a} &= v_{a} \ ,\cr
h_{abc} &= \epsilon_{abcd}v^{d} \ ,\cr
}
\eqn\ansatz
$$
with the other components of $h_{abc}$ vanishing. The reader may
verify that the ansatz respects the self-duality of $h_{abc}$.
The first step in solving the field equations is to calculate the
geometry of the fivebrane for these field configurations. The  matrix
$m$ introduced previously takes the form
$$\eqalign{
m_{\hat a}^{\ \hat  b} &\equiv \delta_{\hat a}^{\ \hat b} -
2h_{\hat a\hat c\hat d}h^{\hat b\hat c\hat d}
\cr &=\left(\matrix{  1+4v^2&0             & \cr
                      0     &1+4v^2        & \cr
                      &     &              &(1-4v^2)\delta_{a}^{\ b}
+8v_{a}v^{b}
\cr }\right) \ .\cr}
\eqn\mdef
$$
where $v^2=v^a v_a$. We can also compute the two metrics $g_{\hat
m \hat n}$ and $G_{\hat m \hat n}$ which occur in the
fivebrane equations of motion \fieldeq\ .  The former is the more
usual brane metric and is given by
$$
g_{\hat m\hat n} =\left(\matrix{
-1&0&\cr
0&1&\cr
&&\delta_{mn} + \partial_{m}\phi\partial_{n}\phi\cr
}\right) \ ,
\eqn\gdef
$$
while the latter metric on the fivebrane becomes
$$\eqalign{
G^{\hat m\hat n} &= \eta^{\hat a\hat b}E_{\hat a}^{\ \hat m}
E_{\hat b}^{\ \hat n}  \cr
&= \left(\matrix{  -(1+4v^2)^2&0             & \cr
                      0       &(1+4v^2)^2    & \cr
                              &              &
(1-4v^2)^2g^{mn} + 16v^{a}v^{b} e_{a}^{\ m}e_{b}^{\ n}\
\cr }\right)   .\cr}
\eqn\Gdef
$$
The vielbein $e_{\hat m}^{\ \hat a}$ associated with
$g_{\hat m\hat n}$
can be found to have the form
$$
e_{\hat m}^{\ \hat a}
=\left(\matrix {  1&0 & \cr
                   0&1 & \cr
                    &  &  \delta_m^a +c\phi_m\phi^a
\cr }\right)\ ,
\eqn\viele
$$
where $\phi_n\equiv \partial_n \phi$,
$c\equiv {({|\phi_n|}^{2})}^{-1}(-1\pm \sqrt{1+|\phi_n|^2})$,
${|\phi_n|}^{2}\equiv \phi_n\phi_m \delta ^{mn}$ and we
adopt the convention that the derivative of $\phi$ only ever carries
a lower world index (i.e. $\phi_a =\delta _a^n \phi_n$). We also find
from equation \gdef\ the following determinants
$$\det g_{mn} = 1+{|\phi_n|}^{2},\ \ 
\det e_{m}^{\ a} = \sqrt{1+{|\phi_n|}^{2}}. \
\eqn\dets
$$
\par
In the calculations
that follow it is important to distinguish  between the two possible
tangent space bases associated with the vielbeins
$e_{\hat a}^{\ \hat m}$ and $E_{\hat a}^{\ \hat m}$ which are in turn 
defined in
terms of  the two metrics
$g_{\hat m \hat n}$ and $G_{\hat m \hat n}$ respectively. We recall that they
are related by
$e_{\hat a}^{\ \hat m} = (m^{-1})_{\hat a}^{\ \hat b}E_{\hat b}^{\ \hat m}$. 
For example $h_{\hat a\hat d\hat e}$ has its indices refered the $E_{\hat
a}^{\ \hat m}$ frame vector fields and is related to the curl of the
gauge field by  [\HSW]
$H_{\hat m\hat n\hat p}=E_{\hat m}^{\ \hat a}E_{\hat n}^{\ \hat
b}E_{\hat m}^{\ \hat c}m_{\hat b}^{\ \hat d}m_{\hat c}^{\ \hat
e}h_{\hat a\hat d\hat e}= e_{\hat m}^{\ \hat a}e_{\hat n}^{\ \hat
b}e_{\hat m}^{\ \hat c}{(m^{-1})}_{\hat c}^{\ \hat
e}h_{\hat a\hat b\hat e}$. We now  calculate the
tensor $H_{\hat m\hat n\hat p}$. The
only non-vanishing components  are
$$\eqalign{
H_{01m} &\equiv V_{m}= {1\over (1+4v^2)}e_m^{\ a}v_a\ ,\cr
H_{mnp}  &= {\det e \over (1-4v^2)} \epsilon_{mnpq}
e^{\ q}_{a}v^{a}\ .\cr}
\eqn\Hdef
$$
\par
Using the equations \Hdef\ and  \viele\ the Bogomol'nyi condition of
equation
\bogomolnyi\ ($V_n=\phi_n$) can be written as
$$
v_a={1\over 2}{\phi_a \over 1+\sqrt{1+ {|\phi_n|}^{2} }}\ .
\eqn\bogtwo
$$
Sustituting the expressions for the determinants of equation
\dets\ we find that the Bogomol'nyi condition of equation
\bogomolnyi\ does indeed coincide with that of equation
\bognon\ . A short calulation using the result of equation \bogtwo\
shows that
$$
H_{mnp}  = \pm {1\over 4}\epsilon_{mnpq} \phi_r \delta ^{rq}\ ,
\eqn\H
$$
and that
$$
G^{mn} =\delta ^{mn}{ 4 \over ({|\phi_n|}^{2})^2}   
{\left( 1\pm \sqrt{1+ {|\phi_n|}^{2}} \right)}^2\propto \delta ^{mn}\ .
\eqn\Gtwo
$$
\par
Let us now  solve the equations of motion \fieldeq . For the ansatz
of equation \ansatz\ the scalar equation becomes
$$G^{mn}\nabla_m\phi_n=0\ ,
\eqn\fieldscalar
$$
while the equation for the gauge field leads to the two equations
$$
G^{mn}\nabla_mV_n=0\ ,
\eqn\fieldgaugeone
$$
and
$$
G^{mn}\nabla_m H_{mpq}=0\ .
\eqn\fieldgaugeone
$$
Clearly, the equation of motion for
$H_{01m}\equiv V_n$ now reduces to that of $\phi$ if we use the
Bogomol'nyi condition.  Using equation \Gtwo\ we find that the scalar
equation \fieldscalar\ becomes
$$\delta^{mn}\partial_{m}\partial_{n}\phi = 0 \ .
\eqn\final
$$
The equation for $H_{mnp}$ becomes an identically satisfied.
\par
We must also ensure that $H_{mnp} $ is closed. We find that the
equation
$$
\epsilon^{mnpq}\partial _m H_{npq}=0\ ,
\eqn\no$$
reduces to the scalar equation \final\ .
\par
Therefore we find the solutions are given by harmonic functions on the flat
transverse space ${\bf R}^4$. For $N$ strings located at $y^{m}_I$, 
$(I=0,1,\ldots,N-1)$
the solution \Hdef\ reduces to
$$\eqalign{
H_{01m} &= \pm{1\over 4}\partial_{m}\phi\ ,\cr
H_{mnp} &=\pm{1\over 4}\epsilon_{mnpq}\delta^{qr}\partial_{r}\phi\ ,\cr
\phi &= \phi_0 + \sum_{I=0}^{N-1} {2Q_i\over \mid x-y_I\mid^2} \ ,\cr }
\eqn\sol
$$
where 
$\phi_0$ and $Q_I$ are constants.
Note that the solution is smooth
everywhere except at the centres of the strings. Furthermore the presence of
the conformal factor in \Gtwo\  implies that the 
equations of motion are
satisfied even at the poles of $\phi$, so that no source terms are required
and the solution is truly solitonic.
Clearly the string soliton \sol\
is self-dual even though
in general the tensor $H_{mnp}$ need not be.
If we
consider a single string
then we find it has the same electric and magnetic charges
$$\eqalign{
Q_E &= {1\over Vol(S_1^3)}\int_{S^3_\infty} \star H = \mp Q_0 \ ,\cr
Q_M &= {1\over Vol(S^3_1)}\int_{S^3_\infty}  H = \mp Q_0\ ,\cr
}\eqn\charges
$$
where $S^3_\infty$ is the transverse sphere at infinity,
$S^3_1$ the unit sphere and $\star$ is the flat six-dimensional Hodge star. 
To calculate the mass per unit length of this string
one could wrap it around a circle of unit circumference and reinterpret the
string as a $0$-brane in five dimensions. The mass per unit length
is then identified with the mass of the $0$-brane. As we will see below
this definition leads to an infinite tension.

Let us now consider the zero modes of a single string soliton, i.e. $N=1$.
Given that we preserve the
asymptotic form of the solution, there are four bosonic zero modes $y_0^{m}$
coming from the  location of the centre of the string. It can be seen that
there are no other bosonic zero modes since these would come from non-zero
expressions for $h_{0mn}$ and $h_{1mn}$ and would ruin the block diagonal 
form of $m_{\hat a}^{\ \hat b}$ which
was crucial for solving the field equations.
This can also be seen from the fact that the solution preserves
eight supercharges and so there is no room for any additional bosonic
zero modes in the two dimensional worldsheet supermultiplet.

The fermionic
zero modes are generated by the broken supersymmetries
$\epsilon_{\alpha}^{\ i}$ which satisfy
$$
\epsilon^{\beta j} = \mp(\gamma^{01})_{\alpha}^{\ \beta}
(\gamma_{ 5'})_i^{\ j}
\epsilon^{\alpha i}.
\eqn\brokensusy
$$
If we call spinors
with $\gamma^{01}\epsilon = \epsilon$ ($\gamma^{01}\epsilon =-\epsilon$)
left (right) moving then we clearly have four left and four right
moving fermion zero modes on the string world sheet, correlated with the
eigenvalue of $\gamma_{ 5'}$.

At first sight the above counting appears to miss
four bosonic zero modes coming from the scalars
$X^{\underline 1},X^{\underline 2},X^{\underline 3},X^{\underline 4}$.
However these scalars do not lead to additional zero modes since they are
fixed by their asymptotic values. This is
analogous to the BPS monopole solutions in $N=4$, $D=4$
super-Yang-Mills where there are six scalars with an $SO(6)$
symmetry relating them. A given BPS monopole will
choose a particular scalar and break this symmetry down to $SO(5)$.
The monopoles obtained using different scalars are to be viewed as distinct,
forming an $SO(6)$ multiplet of monopoles. Similarly, here we obtain a
$SO(5)$ multiplet of self-dual strings.

Thus the string has a $(4,4)$ supermultiplet
of zero modes in agreement with that found in [\DVV]. However,
contrary to our solution,  the strings
described there do not carry any charge with respect to the $H$ field.
Therefore the string soliton found here might represent a D-string in the
six-dimensional self-dual string theory. From the M theory point of view
the strings obtained here should be interpreted as the ends of infinitely
extended membranes. The scalar $X^{5'}=\phi$  then corresponds
to the direction of the membrane transverse to the fivebrane. Clearly the
$SO(5)$ symmetry rotates the choice of this direction. The infinte mass
per unit length can be seen as arising from the infinte length of this
membrane.

%%%%%%%%%%%%%%%%%%%%%%%%%%%%%%%%Chapter four%%%%%%%%%%%%%%%

\chapter{Solitons on D-branes}

It was shown in [\HSW] that M theory fivebrane's equations of motion can
be double dimensionally reduced to the Dirac-Born-Infeld equations of the
type IIA D-fourbrane.
It
follows then that the self-dual string soliton constructed here can also
be reduced to a $0$-brane or $1$-brane solution on the D-fourbrane
worldvolume.
By T-duality it follows that all of the type II D-brane
worldvolume theories admit $0$-brane and  $(p-3)$-brane
solitons preserving half of the
supersymmetry, which can be obtained from the D-fourbrane solutions.
Here we shall content ourselves with the
$p>3$ branes since the lower dimensional branes do not
admit asymptotically free solutions. In what follows below the
background type II supergravity fields are those of flat Minkowski space
and we set all but one of the scalar fields on the D-$p$-brane to be constant.
As with the previous section we use hatted indices to denote all of the
$p+1$ worldvolume coordinates and unhatted indices for the transverse space,
which is $p$ dimensional for $0$-branes or three dimensional for 
$(p-3)$-branes.

First let us wrap the string around the compact dimension
$x^1$ to produce a $0$-brane BPS
soliton on the D-fourbrane worldvolume with the two-form field-strength
$F_{{\hat m}{\hat n}}=H_{{\hat m}{\hat n}1}$ [\HSW]. 
Note that because of the self-duality condition
the other components $H_{{\hat m}{\hat n}{\hat p}}$ do not appear in the  
D-fourbrane's effective
action as an independent field [\HSW]. In this case we obtain
$$\eqalign{
F_{0m}&=\mp{1\over 4}\partial_{m}\phi\ ,\cr
\phi &= \phi_0 + \sum_{I=0}^{N-1} {2Q_I\over \mid x-y_I\mid^2} \ .\cr}
\eqn\zerobrane
$$
We could also dimensionally reduce the self-dual
string soliton by compactifying along $x^5$. In this case we obtain a
string soliton in the D-fourbrane worldvolume. Defining 
$F_{{\hat m}{\hat n}}=H_{{\hat m}{\hat n}5}$ we
obtain
$$\eqalign{
F_{mn}&=\mp{1\over 4}
\epsilon_{mnp}\delta^{pq}\partial_{q}\phi\ ,\cr
\phi &= \phi_0 + \sum_{I=0}^{N-1} {4Q_I\over \mid x-y_I\mid} \ .\cr}
\eqn\onebrane
$$
The generalisation to $p>4$ branes  is now straightforward.
One has $0$-branes with
$$\eqalign{
F_{0m}&=\mp{1\over 4}\partial_{m}\phi\ ,\cr
\phi &= \phi_0 + {4\over p-2}
\sum_{I=0}^{N-1} {Q_I\over \mid x-y_I\mid^{p-2}} \ ,\cr}
\eqn\genzero
$$
and $(p-3)$ branes with
$$\eqalign{
F_{mn}&=\mp{1\over 4}
\epsilon_{mnp}\delta^{pq}\partial_{q}\phi\ ,\cr
\phi &= \phi_0 + \sum_{I=0}^{N-1} {4Q_I\over \mid x-y_I\mid} \ .\cr}
\eqn\genp
$$
The scalar field
only depends on the transverse coordinates and
$\epsilon$ is the flat volume form on the three
dimensional transverse space to the $(p-3)$-brane.
These  solutions preserve half of the worldvolume
supersymmetry and in the case of a single soliton
carry the electric charge $\pm Q_0$ (for \genzero ) or  magnetic charge
$\pm Q_0$ (for \genp ) with respect to $F_{{\hat m}{\hat n}}$.
These solutions represent an infinitely long string or $(p-2)$-brane
respectively,
streching out from the D-$p$-brane along the direction specfied by the
scalar $\phi$.
As with the self-dual string this transverse direction is acted on
by the $SO(9-p)$ symmetry which rotates the scalars.
These solutions are
infinitely massive, similar to a point-like charge in Maxwell's theory. This
is because the Dirac-Born-Infeld expression for the energy does not regulate 
the divergences of the fields at the soliton centres.
However this is  to be expected given the  interpretion as the
end of an infinitely long string or $(p-2)$-brane.

One can also
explicitly verify that
these solitons preserve half the worldvolume supersymmetry at the linearised
order by
compactifying $D=10$ super-Maxwell theory to $p+1$ dimensions, viewed as
the lowest order approximation to the D-brane effective action.
However our construction from the self-dual string solution shows that
\genzero\ and \genp\ are solutions to the non-linear
Dirac-Born-Infeld equations of
motion preserving half of the supersymmetry to all orders.

The D-threebrane is a special case and so we shall treat it separately here.
By wrapping the self-dual string on a torus to four
dimensions  we obtain dyonic
$0$-brane solitons on the D-threebrane worldvolume. Now $x^1$ and $x^5$ are
compact so that  $\phi$ is depends only on $x^{m} = x^2,x^3,x^4$.
Let us consider new coordinates $(u,v)$ related to $(x^1,x^5)$ by
$$
\left(\matrix{u\cr v\cr}\right) = \left(\matrix{a&b\cr c&d\cr}\right)
\left(\matrix{x^1\cr x^5}\right) \ ,
\eqn\newcoords
$$
with $ad-bc=1$.
If we define the two-form field strength 
$F_{{\hat m}{\hat n}}=H_{{\hat m}{\hat n}u}$ then we find
the dyonic solutions
$$\eqalign{
F &= aF^E + c F^M \ ,\cr
\phi &=\phi_0 + \sum_{I=0}^{N-1} {4Q_I\over \mid x-y_I\mid} \ ,\cr}
\eqn\dyons
$$
where $F^E$ and $F^M$ are the purely electric and purely magnetic
field strengths with the components
$$\eqalign{
F^E_{0m} &= \mp{1\over 4} \partial_{m}\phi \ , \cr
F^M_{mn} &= \mp{1\over 4} \varepsilon_{mnp}\delta^{pq}\partial_{q}\phi \ . \cr}
\eqn\Fs
$$
The solution corresponding to a single dyon
then has the electric and magnetic charges
$$
(Q_E,Q_M) = \pm Q_0(a,c) \ .
\eqn\dyoncharges
$$
As is well known the $SL(2,{\bf Z})$ S-duality symmetry of $N=4$, $D=4$
theory acts as
$$
\left(\matrix{a&b\cr c&d\cr}\right)\rightarrow
A^{-1}\left(\matrix{a&b\cr c&d\cr}\right)A\ ,
\eqn\slaction
$$
where $A\in SL(2,{\bf Z})$
and is simply the modular group acting on the torus [\Verlinde,\GLPT].

%%%%%%%%%%%%%%%%%%%%%%%%%%%%%%%%Chapter Five%%%%%%%%%%%%%%%%%%%%%%
\chapter{Discussion}

In this paper we have obtained a self-dual string soliton as a BPS
solution of the M theory fivebrane's equation of motion. Furthermore we
showed that the full non-linear equations of motion and Bogomol'nyi
equation were satisfied to all orders. We also
used the self-dual string soliton to obtain soliton states on type II
D-$p$-branes for $p>2$. Finally we would like to
close with some comments.

It is interesting to note that at the linearised level, where we
the two metrics $G^{mn}$ and $g^{mn}$ are flat and the
tensor $m_a^{\ b}$ is the identity, the field
equations \fieldeq\ become non-interacting. Thus even the theory of a free
tensor multiplet contains a non-trivial BPS soliton in its spectrum. In fact
this statement deserves some qualification since  the theory
isn't completely free at the linearised level
because the fermions still interact with the bosons (although the fermions have
been set to zero in the solution). However,
this does help to clarify why the analysis of the M theory fivebrane's BPS
states carried out in [\DVV] was
so successful, even though it assumed that one could treat the self-dual
string theory as non-interacting.

This unusual situation is what one might expect given the interpretation of
the self-dual string as limit in
which a type IIB D-threebrane wrapped around a 2-cycle of $K3$
becomes light and decouples from gravity [\Witten]. If one considers the
supergravity field equations for a generic $p$-brane there are  equations
for the field strength and  for a scalar field, similar  to \fieldeq\
(with $G^{mn}=g^{mn}$ now identified with the spacetime metric).
\foot{The type IIB threebrane (and hence the related six-dimensional
self-dual string also) is actually unique in this respect
because the dilaton is constant [\DL].}
However, in supergravity what makes these equations interacting and
non-linear is the fact that all these fields couple to the spacetime metric,
even though they don't
couple directly to each other (at least to lowest order in the
supergravity effective action).
Thus in a limit where gravity is decoupling
one may  expect the lowest order bosonic field equations to become those of a
free theory.

While this paper was in the final stages of completion we recieved a
copy of [\CM] and learnt of [\Gibbons]
which overlap with the contents of section three.

\refout

\end